\documentclass[10pt,twocolumn,superscriptaddress,nofootinbib,nobibnotes,secnumarabic,floatfix,aps,prl]{revtex4-2}
\usepackage{placeins}
\usepackage[utf8]{inputenc}
\usepackage[T1]{fontenc}
\usepackage{amssymb,bm,mathrsfs,upgreek}
\usepackage{pifont}
\usepackage[version=4]{mhchem}
\usepackage{dsfont}
\usepackage{graphicx,calligra}
\usepackage{fancyhdr,lastpage}
\usepackage{subcaption,makecell}
\usepackage{ragged2e,caption}
\usepackage{booktabs,array}
\usepackage{soul}
\usepackage{orcidlink}
\usepackage{mathtools}
\usepackage{amsthm}
\usepackage{simpler-wick,slashed}
\usepackage{physics}
\usepackage{hyperref}
\hypersetup{
	colorlinks=true,
	linkcolor=blue,
	urlcolor=blue,
	citecolor=blue
}
\usepackage{cleveref} 

\Crefname{subfigure}{Fig.}{Fig.}
\Crefname{equation}{Eq.}{Eq.}
\Crefname{figure}{Fig.}{Fig.}
\Crefname{section}{Sec.}{Sec.}
\creflabelformat{subfigure}{#2#1#3}

\begin{document}
	\title{Symmetry Breaking as Quantum Gate: Entropy and Weak Mixing Angle}
	
	\author{Qing-Hong Cao}
	\email{qinghongcao@pku.edu.cn}
	\affiliation{School of Physics, Peking University, Beijing 100871, China}
	\affiliation{School of Physics, Zhengzhou University, Zhengzhou 450001, China}
	\affiliation{Center for High Energy Physics, Peking University, Beijing 100871, China}
	
	\author{Yandong Liu}
	\email{ydliu@bnu.edu.cn}
    \affiliation{School of Physics and Astronomy, Beijing Normal University, Beijing, 100875, China}
    \affiliation{Key Laboratory of Multiscale Spin Physics (Ministry of Education), Beijing Normal University, Beijing, 100875, China}
	
	\author{Haotian Qi\,\orcidlink{0009-0004-6989-4134}}
	\email{haotianqi@stu.pku.edu.cn}
	\affiliation{School of Physics, Peking University, Beijing 100871, China}

	\author{Hao Zhang}
	\email{zhanghao@ihep.ac.cn}
	\affiliation{Theoretical Physics Division, Institute of High Energy Physics, Beijing 100049, China}
	\affiliation{School of Physics, University of Chinese Academy of Science, Beijing 100049, China}
	\affiliation{Center for High Energy Physics, Peking University, Beijing 100871, China}
	
	\author{Haoran Zhao}
	\email{haoranzhao@stu.pku.edu.cn}
	\affiliation{School of Physics, Peking University, Beijing 100871, China}

	\begin{abstract}
        We establish a correspondence between two independent entropic probes --- the variation of R\'{e}nyi mutual information (RMI) across the electroweak symmetry breaking (EWSB) transition and the stabilizer Rényi entropy (SRE) --- in tree-level $2\to 2$ elastic scatterings. After angular averaging, the RMI (helicity basis) and the SRE (fixed beam basis) exhibit identical dependence on $\sin^2\theta_W$ within each neutral-current channel. We trace this correspondence to a common physical origin that it's the Yukawa mass insertion acts as a $-\mathrm{i}Y$ quantum gate in chirality space. Minimizing entropies across all processes yields $\sin^2\theta_W$ values matching purely axial vector-like couplings in $Z$ boson exchanged channel. 
	\end{abstract}
	\maketitle

    \textit{Introduction.--} Recently, the interplay between quantum information (QI) and quantum field theory (QFT) has emerged as a compelling perspective for gaining deeper insights into QFT concepts, while quantum simulation (QS) provides a practical platform to probe the intrinsic properties of physical systems~\cite{Bauer:2023qgm,Fang:2024ple}. On one hand, it has been conjectured that, through an extremum condition, the information-theoretic probe of entanglement entropy can be tied to the field-theoretic phenomenon of symmetry breaking~\cite{Cervera-Lierta:2017tdt,Thaler:2024anb,Liu:2025bgw}; on the other hand, by measuring the expectation distribution of operators realized by quantum gates, the entropic diagnostics in QI can be reconstructed from QS.

\begin{figure}[htbp]
        \centering
        \includegraphics[width=\columnwidth]{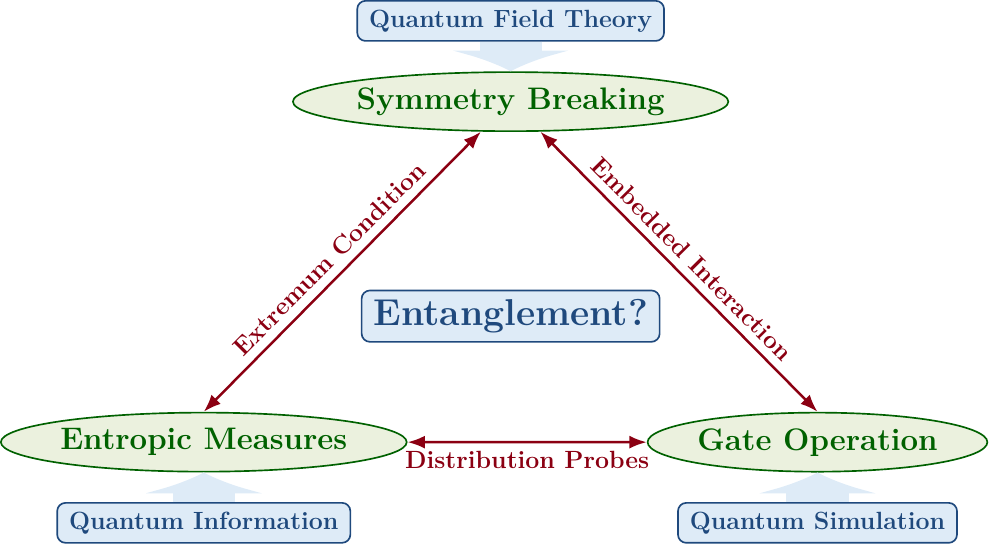}
        \caption{\justifying A unified theoretic framework establishing the correlation between QFT, QI and QS.}
        \label{fig:motivation}
    \end{figure}

    In this work, we demonstrate that the correlation among these three fields is embodied in a specific triad of correspondences, realized by interpreting the Yukawa interactions as quantum gate operations: the gate operation simultaneously characterizes the symmetry breaking and furnishes the operational definition of the entropic diagnostics itself, thereby forming a self-consistent logical loop between QFT, QI and QS, as shown in \Cref{fig:motivation}. The underlying concept of entanglement serves as the common physical origin of all three.

    Specifically, the methodology of the connection between QFT and QI has been applied for comprehending the chirality mixing pattern realized in electroweak symmetry breaking (EWSB). In Ref.~\cite{Cervera-Lierta:2017tdt}, by considering a decay process $Z\to e^+e^-$, $s_W^2=1/4$ was discovered analytically via maximization of concurrence in the helicity basis. In addition, based on minimizing 2-order stabilizer R\'{e}nyi entropy (SRE) in the M\o{}ller scattering process, Ref.~\cite{Liu:2025bgw} discovered the association between magic minimization and the weak mixing angle.

    We introduce the quantum R\'{e}nyi mutual information (RMI) \cite{Berta:2014vma} as an independent measure to quantify quantum correlations between bi-partite subsystems
     \begin{equation}
        I_\alpha(A:B)=S_\alpha(\rho_A)+S_\alpha(\rho_B)-S_\alpha(\rho_{AB})
         \label{eq:RMI}
     \end{equation} 
    where $S_\alpha(\rho)=(1-\alpha)^{-1}\log(\Tr(\rho^\alpha))$ is the $\alpha-$order R\'{e}nyi entropy, and $\rho_{A}\coloneqq\Tr_B(\rho_{AB}),\rho_{B}\coloneqq\Tr_A(\rho_{AB})$ are the reduced density matrices of subsystems $A$ and $B$.
    
    The variation in quantum correlation between the two physical scenarios, before and after EWSB, reflects a preference for the orientation of EWSB, as captured by the variation in RMI across the EWSB transition. In the symmetric phase, the degrees of freedom are massless, with dynamics dictated by the $SU(2)_L \times U(1)_Y$ gauge couplings $g_2$ and $g_Y$. Upon spontaneous symmetry breaking, the fermions and the weak gauge bosons ($W^\pm, Z$) acquire mass, while the photon remains massless; the resulting vacuum orientation within the weak isospin and hypercharge space is characterized by the weak mixing angle $\theta_W$, satisfying the parameterization $s^2_W=g_Y^2/(g_Y^2+g_2^2)$.
    
\begin{figure}[ht!]
        \centering
        \includegraphics[width=\columnwidth]{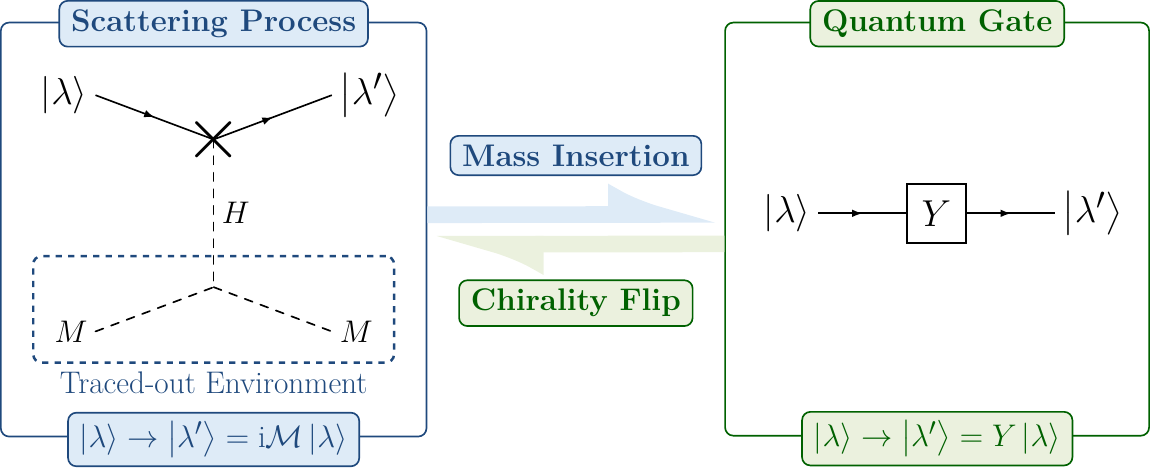}
        \caption{\justifying Mass insertion as scattering process mediated by Higgs is equivalent to quantum gate operating in the chirality basis.}
        \label{fig:massinsertion}
    \end{figure}

\begin{figure}[ht!]
        \centering
        \includegraphics[width=0.9\columnwidth]{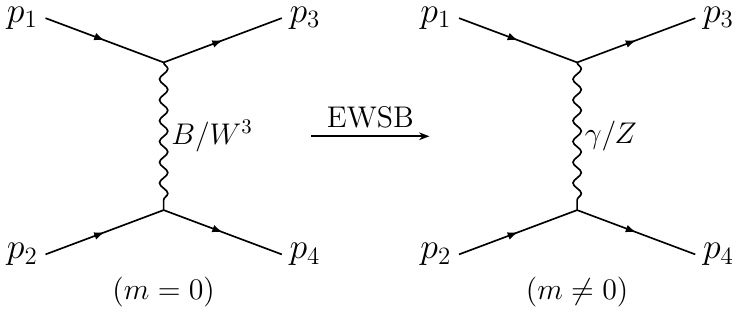}
        \caption{\justifying Scattering process mediated by different neutral current before and after the EWSB. The external legs of fermions also acquire finite mass after EWSB.}
        \label{fig:feynman diagrams}
    \end{figure}

    To reveal the $s^2_W$ dependence in the symmetric and broken phases, we investigate the evolution of final state density matrices in the helicity degree of freedom reconstructed by $2 \to 2$ elastic scattering processes across the electroweak phase transition, denoted as $\rho^{(\mathrm{S})}$ in the symmetric phase and $\rho^{(\mathrm{B})}$ in the broken phase. Direct evaluation shows that in the high energy limit, the variation of 2-order RMI between $\rho^{(\mathrm{S})}$ and $\rho^{(\mathrm{B})}$, i.e.,
        \begin{equation}
    \Delta_{I}\big(\rho^{(\mathrm{B})},\rho^{(\mathrm{S})}\big)
      = \abs{I_2^{(\mathrm{B})}(A:B)-I_2^{(\mathrm{S})}(A:B)},
     \label{eq:REdef}
     \end{equation}
     reproduces the same dependence on $s_W^2$ of as the SRE. As we demonstrate below (\Cref{fig:massinsertion,fig:feynman diagrams}), this agreement holds quantitatively after angular averaging and across multiple fermion channels.

    \textit{Yukawa Interaction as Quantum Gate.--} The parameter consistency between SRE and RMI in helicity space can be viewed as an underlying correspondence between the interaction realizing symmetry breaking and a quantum gate. From the perspective of RMI, the dominant contribution to $\Delta_{I}\big(\rho^{(\mathrm{B})},\rho^{(\mathrm{S})}\big)$ arises from the Yukawa interaction $\mathcal{L}_{\text{Yukawa}}=-m\left(\bar{\psi}_L\psi_R+\bar{\psi}_R\psi_L\right)$ across the symmetric ($m=0$) and broken ($m\neq0$) phases. Meanwhile, the finite mass insertion is an operation along with the same fermionic line, and can therefore be evaluated as a scattering process in the chirality basis. Maintaining the $2\to2$ kinematics, we consider the $f+M\to f+M$ process mediated by the Higgs boson, where the heavy scalar $M$ is treated as a 1-dimensional traced-out environment in chirality space. 
    
    The scattering amplitudes $\mathcal{M}_{\lambda'\lambda}\propto\braket{\lambda'}{\lambda}$ are given by:
    \begin{equation}
        \label{eq:Ygate}\mathcal{M}_{\lambda'\lambda}\propto\begin{pmatrix}
            0 & -\left(1-\frac{M^2}{s}\right)\sin\frac{\theta}{2}\\
            \left(1-\frac{M^2}{s}\right)\sin\frac{\theta}{2} & 0
        \end{pmatrix}\propto -\mathrm{i}Y
    \end{equation}
    This is equivalent to the $-\mathrm{i}Y$ gate ($Y=\sigma^2$) after normalization in the orthonormal computational chirality basis, spanned by the spinors $|\lambda_\pm \rangle$ of fermions in the Weyl representation, 
	 \begin{equation}\label{helicitybasis1} 
	 		\ket{\lambda_+}=\begin{pmatrix}			\omega_+\ket{\hat{\bm{p}},+}\\
	 			\omega_{-}\ket{\hat{\bm{p}},-}
	 		\end{pmatrix},\qquad 
	 		\ket{\lambda_-}=\begin{pmatrix}
	 			\omega_+\ket{\hat{\bm{p}},-}\\
	 			\omega_-\ket{\hat{\bm{p}},+}
	 		\end{pmatrix}
	 \end{equation}
    with $\omega_{\pm}=\sqrt{E\pm\left|\bm{p}\right|}$, where $E$ and $\bm{p}$ are the energy and momentum of the fermions, respectively;       
	 \begin{equation}\label{helicitybasis2}
	 	\begin{gathered}
	 		\ket{\hat{\bm{p}},+}=\left(\begin{array}{cc}
	 			\cos\frac{\theta}{2}\\
	 			e^{\mathrm{i}\phi}\sin\frac{\theta}{2}
			\end{array}\right),~
	 		\ket{\hat{\bm{p}},-}=\left(\begin{array}{cc}
	 			-e^{-\mathrm{i}\phi}\sin\frac{\theta}{2}\\
	 			\cos\frac{\theta}{2}
	 		\end{array}\right)
	 	\end{gathered}
	 \end{equation}
	 with $\hat{\bm{p}}=\left(\sin\theta\cos\phi,\sin\theta\sin\phi,\cos\theta\right)$ describing the orientation of the 3-momentum in the center-of-mass (c.m.) frame determined by scattering kinematics $(\theta,\phi)$.

    \textit{Application: $e^{-} \mu^{-} \rightarrow e^{-} \mu^{-}$.--} To verify the gate interpretation quantitatively, we evaluate the density matrices for a benchmark process. Taking the $e^{-} \mu^{-} \rightarrow e^{-} \mu^{-}$ scattering process, we present an explicit evaluation of the density matrix characterizing the EWSB transition. As shown in \Cref{fig:feynman diagrams}, the final-state density matrices are determined by performing a coherent summation over scattering amplitudes. We employ the helicity amplitude formalism to evaluate the 16 distinct helicity configurations contributing to the process.

    Working in the orthonormal basis of helicity eigenstates $\ket{\pm}$, we adopt a maximally mixed initial state $\rho^{(\text{in})}=\mathds{1}_{4\times4}/4$ for the computation of RMI. This choice naturally mimics an unpolarized beam or a state that has decohered in a thermal bath. Comparatively, for the evaluation of SRE (defined for pure states), we use the Haar measure over the Clifford group and select 60 stabilizer states $\ket{\psi_{\text{stab}}^{(i)}}$ with equal weights, each as initial states respectively, which form a spherical 2‑design over the full two‑qubit Hilbert space. The two alternative choices of initial states are consistent in the first moment, i.e.,
    \begin{equation}\label{eq:initial}
        \frac{1}{60}\sum_{i=1}^{60}\ket{\psi_{\text{stab}}^{(i)}}\bra{\psi_{\text{stab}}^{(i)}}=\mathds{1}_{4\times 4}/4
    \end{equation}

	Given the initial state $\rho^{(\text{in})}$, quantum state tomography reconstructs the final state density matrix $\rho^{(\text{out})}$ from the scattering process, 
	\begin{equation}\label{eq:finaldensitymatrix}
		\rho^{(\text{out})}
        =\frac{\mathcal{M}\rho^{(\text{in})}\mathcal{M}^\dagger}{\Tr(\mathcal{M}\rho^{(\text{in})}\mathcal{M}^\dagger)}
	\end{equation}
where $\mathcal{M}$ is the helicity amplitude for M\o{}ller scattering in the computational basis, 
	\begin{equation}\label{eq:helicityamplitude}
		\mathcal{M}_{h_1h_2h_3h_4}=\frac{g_V^2}{q^2-m_V^2} \mel{\lambda_{h_3}}{\Gamma^{\mu}}{\lambda_{h_1}}\mel{\lambda_{h_4}}{\Gamma_{\mu}}{\lambda_{h_2}}
	\end{equation}
	where $g_V$ is the overall coupling strength, $q$ is the transferred momentum and $m_V$ is the pole mass of the vector boson. $h_i$ denotes the helicity of the $i$-th external particle, and
	\begin{equation}\label{eq:chiralvertice}
		\Gamma^\mu=\gamma^\mu\left(\kappa_L P_L+\kappa_R P_R\right)
	\end{equation}
 	with $P_{R,L}=\left(1\pm\gamma^5\right)/2$ being the chirality projector, and $\kappa_{R,L}$ denoting the relative coupling strength to the right- and left-handed fermions; see \Cref{tab:fermionspectrum}.     

    \begin{table}[htbp]
        \caption{\justifying The SM fermion spectrum carrying gauge charges of $SU(2)_L$, $U(1)_Y$ and $U(1)_{\text{em}}$, in which $Y_W$, $T^3_W$ and $Q$ denote the weak hypercharge, the weak isospin and the electric charge, respectively. We highlight the representative $\kappa_{L,R}^{(Z)}$ for the $Z$-exchanged channel due to its non-trivial dependence on the mixing angle $s_W^2$.}
	\label{tab:fermionspectrum}
	\centering
	\resizebox{\linewidth}{!}{\begin{tabular}{cccccccc}
								\toprule[0.5mm]
								Flavor & $Y^W_L$ & $Y^W_R$ & $T^{W,3}_L$ & $T^{W,3}_R$ & $Q$ & $\kappa_L^{(Z)}$ & $\kappa_R^{(Z)}$ \\
								\midrule
								$u,c,t$ & $+\frac{1}{3}$ & $+\frac{4}{3}$ & $+\frac{1}{2}$ & 0 & $+\frac{2}{3}$ & $\frac{1}{2}-\frac{2}{3}s_W^2$\quad & $-\frac{2}{3}s_W^2$ \\
								\addlinespace[1em]
								$d,s,b$ & $+\frac{1}{3}$ & $-\frac{2}{3}$ & $-\frac{1}{2}$ & 0 & $-\frac{1}{3}$ & $\frac{1}{3}s_W^2-\frac{1}{2}$ & $+\frac{1}{3}s_W^2$ \\
								\addlinespace[1em]
								$e,\mu,\tau$ & $-1$ & $-2$ & $-\frac{1}{2}$ & 0 & $-1$ & $s_W^2-\frac{1}{2}$ & $+s_W^2$ \\
								\addlinespace[1em]
								$\nu_e,\nu_\mu,\nu_\tau$ & $-1$ & $\setminus$ & $+\frac{1}{2}$ & $\setminus$ & $0$ & $+\frac{1}{2}$ & $\setminus$ \\
								\bottomrule[0.5mm]
						\end{tabular}}
    \end{table}
    
    In the high-energy limit, the two theories before and after EWSB coincide, and the leading non-vanishing contribution to the RMI scales as
    \begin{equation}\label{orderexpansion}
        \Delta_I\big(\rho^{(\mathrm{B})},\rho^{(\mathrm{S})}\big)=\left(\frac{m^2}{s}\right)\times\Delta_\mathcal{I}(\rho^{\mathrm{(B)}},\rho^{\mathrm{(S)}})
    \end{equation}
    where $\Delta_\mathcal{I}(\rho^{\mathrm{(B)}},\rho^{\mathrm{(S)}})(\theta, s_W^2)$ is an energy-independent function of the scattering angle and weak mixing angle. In the following, we focus exclusively on this dimensionless contribution $\Delta_\mathcal{I}(\rho^{\mathrm{(B)}},\rho^{\mathrm{(S)}}) $.
	
    In the  
    $e^-\mu^- \to e^-\mu^-$ scattering process, we compute the $\Delta_\mathcal{I}(\rho^{\mathrm{(B)}},\rho^{\mathrm{(S)}})$ in the helicity basis and compare our results with the 2-order SRE in the fixed beam basis with 60 initial stabilizer states averaged $\overline{\mathcal{M}_2}$ studied in Ref.~\cite{Liu:2025qfl,Liu:2025bgw}. 
    This comparison provides a direct consistency check between two independent information-theoretic probes, each with a different initial state and computational basis. In the numerical evaluation of $\overline{\mathcal{M}_2}$, we adopt a c.m. energy of $\sqrt{s}=10~\mathrm{TeV}$, following Ref.~\cite{Liu:2025bgw}, as representative of the high-energy regime.
    
\begin{figure}[htbp]
		\centering
		\includegraphics[width=\columnwidth]{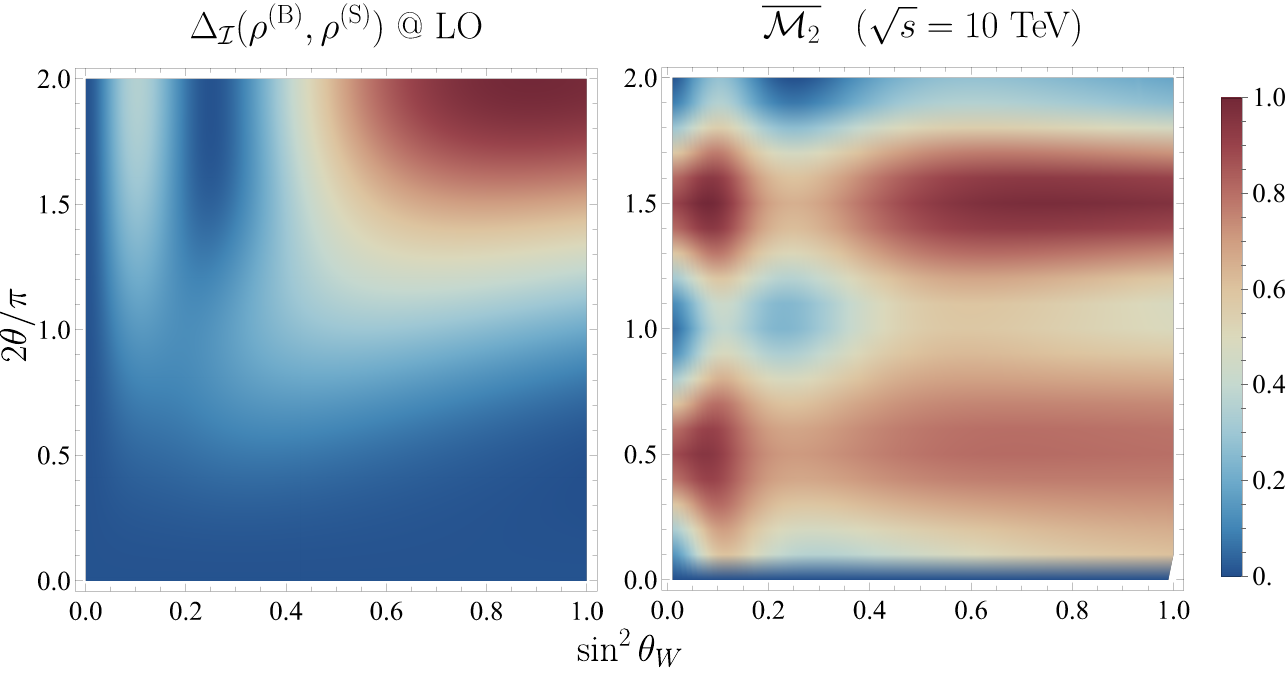}
		\caption{\justifying RMI $\Delta_\mathcal{I}\left(\rho^{(\mathrm{B})},\rho^{(\mathrm{S})}\right)$ (left panel) and SRE $\overline{\mathcal{M}_2}$ (right panel) in $e^-\mu^- \to e^-\mu^-$ scattering, showing their dependence on scattering kinematics and the weak mixing angle $(\theta,\sin^2\theta_W)$. The quantities $\Delta_\mathcal{I}\left(\rho^{(\mathrm{B})},\rho^{(\mathrm{S})}\right)$ and $\overline{\mathcal{M}_2}$ are normalized to $[0,1]$ respectively.}
		\label{RE/SRE(eueu)}
	\end{figure}

    The dependence of the final state $\Delta_\mathcal{I}\left(\rho^{(\mathrm{B})},\rho^{(\mathrm{S})}\right)$ and $\overline{\mathcal{M}_2}$ on the scattering angle $\theta$ and the weak mixing angle $s_W^2$ are shown in \Cref{RE/SRE(eueu)}. 
    A salient feature visible in \Cref{RE/SRE(eueu)} is that RMI exhibits a qualitatively different dependence on kinematics compared with SRE. 
    For SRE constructed in the spin computational basis, Refs.~\cite{Thaler:2024anb,Martin:2025hzm} show that perpendicular scattering ($\theta=\pi/2$) plays a privileged role, as it does in the concurrence. 
    In contrast, the $\Delta_\mathcal{I}\left(\rho^{(\mathrm{B})},\rho^{(\mathrm{S})}\right)$ does not favor any particular scattering orientation. 
    As discussed in~\cite{Peres:2002wx,He:2007gk}, this distinction originates from the global Lorentz rotation relating the helicity and the fixed beam computational bases. 
    
    Another notable feature is the emergence of a decoherence-like behavior in the forward-scattering region. 
    As seen in \Cref{RE/SRE(eueu)}, RMI becomes increasingly insensitive to $s_W^2$ as $\theta \to 0$. 
    This behavior can be understood as a consequence of the strong enhancement of massless gauge-boson exchange in the forward limit, which effectively drives the final-state density matrix toward the maximally mixed state, thereby suppressing its sensitivity to the electroweak parameter. 
    Inspired by recent analyses of soft-radiation effects~\cite{Gu:2025ijz}, this feature provides an independent consistency check of our computational framework, given our prior choice of $\rho^{(\text{in})}\propto \mathds{1}$. 

    In order to compare with the SRE result without interferences from scattering kinematics, we average the RMI over the celestial sphere with uniform weight:
    \begin{equation}
    \langle \Delta_\mathcal{I} \rangle = \frac{1}{4\pi}\int_{\Omega} \dd{\Omega} \, \Delta_\mathcal{I}(\theta,\phi).
    \end{equation}
    
\begin{figure}[htbp]
		\centering
		\includegraphics[width=\columnwidth]{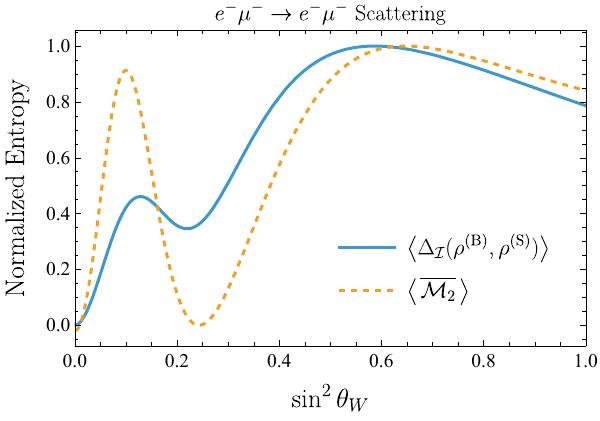}
		\caption{\justifying Angular averaged RMI $\expval{\Delta_\mathcal{I}\left(\rho^{(\mathrm{B})},\rho^{(\mathrm{S})}\right)}$ (solid blue) and angular averaged SRE $\expval{\ \!\overline{\mathcal{M}_2}\ \!}$ (dashed orange) in $e^-\mu^- \to e^-\mu^-$ scattering as a function of $\sin^2\theta_W$.
        The quantities $\expval{\Delta_\mathcal{I}\left(\rho^{(\mathrm{B})},\rho^{(\mathrm{S})}\right)}$ and $\expval{\ \!\overline{\mathcal{M}_2}\ \!}$ are normalized to $[0,1]$ respectively.}
		\label{fig:RE/SRE(eueu)average}
	\end{figure}

    After applying the angular-averaging prescription, the RMI and SRE exhibit a consistent dependence on the weak mixing angle. As shown in \Cref{fig:RE/SRE(eueu)average}, both $\expval{\Delta_\mathcal{I}\left(\rho^{(\mathrm{B})},\rho^{(\mathrm{S})}\right)}$ and $\expval{\ \!\overline{\mathcal{M}_2}\ \!}$ retain consistent dependence on the weak mixing angle $s_W^2$, with the minimization point at $\left(s_W^2\right)_{\text{min}}\approx1/4$.

    \textit{Origin of the RMI-SRE Correspondence: $\mathbb{Z}_2$ Reduced SRE.--} The angular-averaged comparison demonstrates that RMI and SRE agree numerically; we now investigate the underlying physical principle by decomposing SRE into its Pauli-string components.

    To trace back the origin of the qualitative consistency between RMI and SRE and the common emergence of $\left(s_W^2\right)_{\text{min}}\approx1/4$, we consider a $\mathbb{Z}_2$ reduction form of 2-order SRE, denoted by $\widetilde{\mathcal{M}}_2$. The original concept of \(\alpha-\)order SRE for an $n$-qubit system is the \(\alpha\)-R\'{e}nyi entropy associated with a complete set of probability distribution on $SU(2)^{\otimes n}$ space, explicitly written as the squared (normalized) expectation value of \(\mathcal{P}\) in the pure state \(\ket{\psi}\), $\Xi_\mathcal{P}(\ket{\psi}) := d^{-1} \abs{\mel{\psi}{\mathcal{P}}{\psi}}^2$, where $d=2^n$ is the dimension of the total Hilbert space. The completeness of the probing distribution is realized by the Pauli string operators $\mathcal{P}\in\mathcal{P}_n$ up to phases \(\{\pm 1,\pm\mathrm{i}\}\):
    \begin{equation}
        \mathcal{P}_n=P_1\otimes P_2\otimes\cdots \otimes P_n, \quad P_i\in\{I,X,Y,Z\}
        \label{eq:Paulistring}
    \end{equation}
    where $I$ is the identity gate, and $X=\sigma^1,Y=\sigma^2,Z=\sigma^3$ are the three Pauli matrices defined along three fixed orthogonal axes. Alternatively, the $\mathbb{Z}_2$ reduced SRE is supported on a closed subset of measurements $\{I,X\},\ \{I,Y\}$ or $\{I,Z\}$ (up to phases $\{\pm1,\pm\mathrm{i}\}$) for each particle in the Pauli string measurement, forming a reduced distribution and similarly characterized by the 2-order R\'{e}nyi entropy.
    
\begin{figure}[htbp]
     \centering
     \begin{subfigure}[b]{.96\columnwidth}
        \includegraphics[width=\columnwidth]{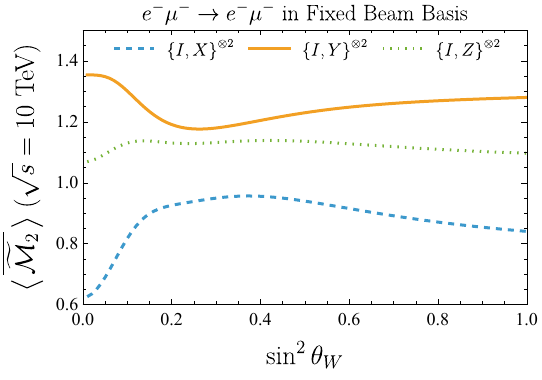}
     \end{subfigure}
     \begin{subfigure}[b]{.96\columnwidth}
        \includegraphics[width=\columnwidth]{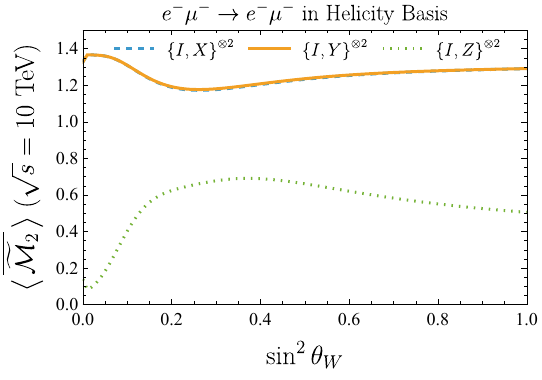}
     \end{subfigure}
     \caption{\justifying $\mathbb{Z}_2$ reduced SRE in fixed beam (above) and helicity (below) basis. The selection on $\mathbb{Z}_2$ reduced string operators $\mathcal{P}_i\in\{I,X\}$ (dashed blue), $\mathcal{P}_i\in\{I,Y\}$ (solid orange) and $\mathcal{P}_i\in\{I,Z\}$ (dotted green) are plotted respectively.}
     \label{fig:reducedSRE}
 \end{figure}

    It should be emphasized that the Pauli string measurement is basis dependent, in which three Pauli gates are related to each others under coordinate transformations. As shown in \Cref{fig:reducedSRE}, we compute the scattering angle averaged $\big<\overline{\widetilde{\mathcal{M}}_2}\big>$ with the same 60 stabilizer initial states summed over under the fixed beam basis and the helicity basis respectively.
 
    It turns out that under both bases, the dominant channel for the emergence of $\left(s_W^2\right)_{\text{min}}\approx1/4$ is $\{I,Y\}$, validating the physical intuition from \Cref{eq:Ygate}. From the perspective of trace-type entropic measures such as RMI, these capture the diagonal terms of the density matrices. Consequently, the leading contribution to the variation of RMI at $\mathcal{O}(m^2/s)$ originates from twice chirality flips realized as twice mass insertions, which can be factorized from the overall scattering process and further characterized by the  $-\mathrm{i}Y$ gate operation illustrated in \Cref{fig:massinsertion}.

    The logical chain is therefore: (i) the leading mass correction enters as two chirality flips along the fermion line; (ii) each flip acts as a $-\mathrm{i}Y$ gate in the computational chirality basis; (iii) the $\{I,Y\}$ Pauli-string subset of SRE is the only component sensitive to this gate; (iv) the resulting $s_W^2$ dependence is shared between the diagonal (trace-type) contribution captured by RMI and the $\{I,Y\}$ projection captured by SRE.
     
    Therefore, we conclude that the non-vanishing Yukawa interaction that generating EWSB acts as a quantum gate and thereby leads to the same dependence on $s_W^2$ for both entropic measures, SRE and RMI.

    \textit{Extension to Multiple Fermion Channels.--} To assess whether the extremal value $\left(s_W^2\right)_{\text{min}}\approx1/4$ reflects an underlying physical principle or merely arises as a channel-dependent numerical coincidence, we extend our analysis to a broader class of neutral-current–mediated scattering processes involving different fermion species, carrying distinct electroweak charges listed in \Cref{tab:fermionspectrum}. Applying the same angular-averaging prescription, we evaluate the RMI and SRE for the processes $e^-\mu^- \to e^-\mu^-$, $uc \to uc$, and $ds \to ds$, while neutrino channels are omitted since no neutral-current chiral mixing is present. 
    \begin{figure}[htbp]
	\centering
        \begin{subfigure}[b]{.96\columnwidth}
            \includegraphics[width=\columnwidth]{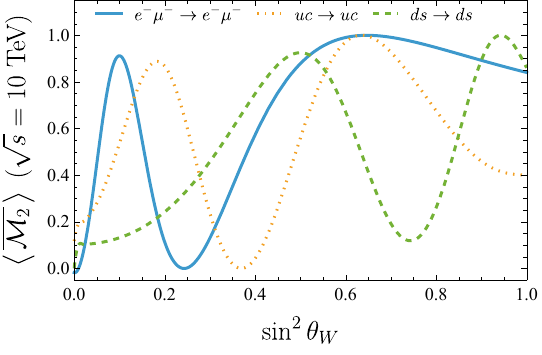}
        \end{subfigure}
	\begin{subfigure}[b]{.96\columnwidth}
            \includegraphics[width=\columnwidth]{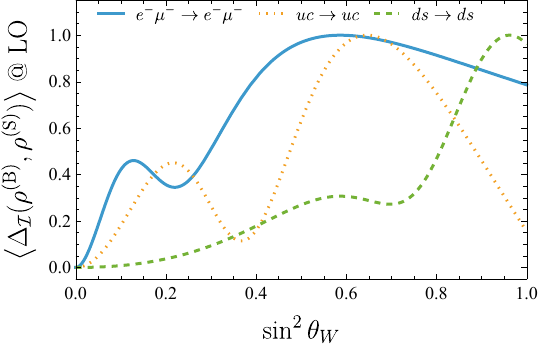}
        \end{subfigure}
	\caption{\justifying The normalized angular averaged SRE (above) and RMI (below) for $e^-\mu^-\to e^-\mu^-$ (solid blue), $uc\to uc$ (dotted orange) and $ds\to ds$ (dashed green) scattering processes respectively.}
	\label{SREandRE}
    \end{figure}
    
    The resulting angular-averaged entropies $\expval{\ \!\overline{\mathcal{M}_2}\ \!}$ and $\expval{\Delta_\mathcal{I}\left(\rho^{(\mathrm{B})},\rho^{(\mathrm{S})}\right)}$, shown in \Cref{SREandRE}, reveal that the minimization point $\left(s_W^2\right)_{\text{min}}$ is not universal across fermion species. In particular, extending the analysis beyond leptonic channels demonstrates that the value $\left(s_W^2\right)_{\text{min}}\approx1/4$ does not persist in general, indicating that its appearance in specific processes is accidental rather than fundamental. This behavior is analogous to the finding reported in \cite{Thaler:2024anb}, highlighting that flavor structure encoded in the helicity amplitudes generates different entropic behavior, leading to different extremal points of $s_W^2$.
    \begin{table}[ht!]
		\centering
            \caption{\justifying $(s_W^2)_{\text{min}}$ value predicted by entropy minimization among different fermion species.}
		\begin{tabular*}{\columnwidth}{@{\extracolsep{\fill}}cccc@{}}
			\toprule[0.5mm]
			& $e^-\mu^-\to e^-\mu^-$\hspace{1ex} & $uc\to uc$\hspace{1ex} & $ds\to ds$\hspace{1ex} \\
			\midrule
			$(s_W^2)_{\text{min}}\text{ from RMI}$ & 0.23 & 0.37 & 0.72 \\[1em]
			$(s_W^2)_{\text{min}}\text{ from SRE}$ & 0.24 & 0.37 & 0.74 \\
			\bottomrule[0.5mm]
		\end{tabular*}
		\label{tab:minimalpoint}
	\end{table}
    
    More precisely, the location of the entropy minimization is controlled by the chiral structure of the underlying interaction. The numeral values shown in \Cref{tab:minimalpoint} implies that only in $Z$-exchanged channels with purely axial vector-like couplings, characterized by $\kappa_L^{(Z)}=-\kappa_R^{(Z)}$, does the extremum occur near $\left(s_W^2\right)_{\text{min}}\approx1/4$. For fermions with asymmetric left- and right-handed couplings, the minimization point is shifted accordingly. This analysis clarifies that the role of RMI (and SRE) is not to predict a universal value of the weak mixing angle, but rather to diagnose the chiral structure encoded in the scattering amplitude.

    \textit{Conclusion.--} In this Letter, we have demonstrated that the R\'{e}nyi mutual information variation across the EWSB transition and the stabilizer R\'{e}nyi entropy, despite being defined in different bases with different initial states, yield identical dependence on $s_W^2$ in tree level neutral-current $2\to2$ scatterings. We traced this correspondence to a common origin: the Yukawa mass insertion acts as a $-\mathrm{i}Y$ gate in chirality space, and its double application generates the leading $\mathcal{O}(m^2/s)$ contribution to RMI and the dominant reduced channel $\{I,Y\}$ in SRE. The entropy minimum in each channel diagnoses the axial chiral coupling structure $\kappa_R^{(Z)}\approx-\kappa_L^{(Z)}$ rather than predicting a universal value of $s_W^2$. This triad that symmetry breaking as gate operation and entropic diagnostics establishes a self-consistent loop connecting QFT, QI, and QS.

    Bias tendency on angular orientation serves as flexible choice of computational basis utilized for optimally detecting maximum entanglement in real world collider experiments \cite{Cheng:2024btk}. However, ambiguity arises from the basis transformation along with the specific gate operation. It is still an open question to decouple the classical correlation induced by scattering kinematics from the quantum entanglement induced by intrinsic parameters. 

    Another important open question remains the construction of a state-independent entropy measure for scattering processes, which would remove the current sensitivity to initial-state assumptions \cite{Low:2024hvn,Chang:2024wrx}. More broadly, whether density-matrix tomography in high-energy scattering can extract information beyond the elastic cross section \cite{Seki:2014cgq,Peschanski:2016hgk,Low:2024hvn} remains a compelling direction for future work.
	
	\textit{Acknowledgment.--} We would like to thank Zhewei Yin, Ding Yu Shao for useful discussions. This work is partially supported by the National Natural Science Foundation of China under Grant Nos. 12235001, 12075257, 12175016 and 12575114. The work of Y. L. is also partly supported by the National Key R\&D Program of China under Grant No. 2023YFA1607104, and Fundamental Research Funds for the Central Universities, Beijing Normal University. H. Q. is also supported by Beijing Natural Science Foundation under Grant No. QY26015. This work was also supported by the Fundamental and Interdisciplinary Disciplines Breakthrough Plan of the Ministry of Education of China-JYB2025XDXM204. The authors gratefully acknowledge the valuable discussions and in-sights provided by the members of the Collaboration on Precision Tests and New Physics (CPTNP).
    \bibliography{refs}
    
\end{document}